\documentclass[final,3p,times]{elsarticle}

\biboptions{sort&compress}

\usepackage{amssymb}
\usepackage{amsmath}

\usepackage{graphicx}

\usepackage[HTML]{xcolor}
\definecolor{my_magenta_color}{HTML}{785EF0}
\definecolor{my_pink_color}{HTML}{DC267F}
\definecolor{my_yellow_color}{HTML}{FFB000}

\newcommand{\Phiave}{\langle{\Phi}\rangle}
\newcommand{\Phirms}{\Phi_{\text{rms}}}
\newcommand{\Phitilde}{\widetilde{\Phi}}

\newcommand{\Figref}[1]{Fig.~\ref{#1}}

\begin{document}

\begin{frontmatter}



\title{Blob velocities and sizes in the Alcator C-Mod scrape-off layer for ohmic and high confinement mode plasmas}


\author[UiT]{A.~D.~Helgeland}

\author[UiT]{J.~M.~Losada}

\author[PSFC]{J.~L.~Terry}

\author[UiT]{O.~E.~Garcia}

\affiliation[UiT]{
    organization={UiT The Arctic University of Norway},
    addressline={Department of Physics and Technology},
    city={Troms{\o}},
    postcode={N-9037},
    country={Norway}}

\affiliation[PSFC]{
    organization={MIT Plasma Science and Fusion Center},
    city={Cambridge},
    postcode={02139},
    state={Massachusetts},
    country={USA}
}

\begin{abstract}
An improved time delay estimation method is used to calculate the velocity of cross-field blob motion in the scrape-off layer of Alcator C-Mod for an ohmic and two high confinement (H-mode) plasmas; an edge localized mode free and an enhanced D-alpha H-mode. The gas puff imaging data analysis results are interpreted in the framework of a stochastic model that describes the fluctuations as a super-position of uncorrelated blob-like structures. In all confinement modes investigated, the scrape-off layer is dominated by large amplitude, blob-like filaments moving radially outwards with velocities in the range from $400$ to $1000\,$m/s. Blobs in high confinement mode plasmas have similar velocities and sizes as in ohmic plasma, which is consistent with the close similarity of conditionally averaged burst shapes and frequency spectra for the confinement modes investigated.
\end{abstract}




\end{frontmatter}


\section{Introduction}
\label{sec:intro}

Cross-field transport of particles and heat in the boundary region of magnetically confined plasmas is dominated by radial motion of blob-like filament structures \cite{dippolito_convective_2011}. Their motion results in large-amplitude and intermittent fluctuations in the scrape-off layer (SOL), as well as flattened time-average radial profiles, both which leads to enhanced plasma interactions with limiter structures and the main chamber wall \cite{labombard_particle_2001,terry_scrape-off_2007}. This will impact the lifetime of plasma facing components at the outboard mid-plane region in the next generation, high-duty cycle magnetic confinement experiments and future fusion reactors \cite{pitts_material_2005,lipschultz_plasma-surface_2007}. It is therefore important to clarify the properties of plasma fluctuations and blob-like filaments in the SOL.

Previous investigations on Alcator C-Mod have demonstrated a remarkable similarity of the fluctuation statistics between ohmic and high confinement mode (H-mode) plasmas \cite{terry_transport_2005,labombard_new_2014,kube_fluctuation_2016,garcia_burst_2013,garcia_intermittent_2013,garcia_intermittent_2018,theodorsen_universality_2018,kube_intermittent_2018,kube_comparison_2020,ahmed_strongly_2023,theodorsen_relationship_2017,kube_statistical_2019,kuang_plasma_2019}. In particular, the large-amplitude bursts recorded by the gas puff imaging (GPI) diagnostic in the SOL have an exponential waveform and exponentially distributed amplitudes and waiting times \cite{garcia_burst_2013,garcia_intermittent_2013,kube_fluctuation_2016,garcia_intermittent_2018,theodorsen_universality_2018,kube_intermittent_2018,kube_comparison_2020,ahmed_strongly_2023}. Moreover, the probability density functions (PDFs) and frequency power spectral densities (PSDs) are similar for all these confinement modes \cite{garcia_burst_2013,garcia_intermittent_2013,theodorsen_relationship_2017,garcia_intermittent_2018,theodorsen_universality_2018,kube_intermittent_2018,kube_comparison_2020,ahmed_strongly_2023}.

The PDFs and PSDs are in excellent agreement with a stochastic model that describes the fluctuations as a super-position of uncorrelated pulses \cite{garcia_stochastic_2012,theodorsen_probability_2018,garcia_stochastic_2016,militello_scrape_2016,militello_relation_2016,losada_stochastic_2023,losada_stochastic_2024,garcia_auto-correlation_2017}. According to this statistical framework, the plasma fluctuations are Gamma distributed with the shape parameter given by the ratio of average pulse duration and waiting times, and the scale parameter given by the average pulse amplitude \cite{garcia_stochastic_2012,garcia_stochastic_2016,theodorsen_probability_2018}. The PSD is determined by the pulse shape, and is the product of two Lorentzian spectra for a two-sided exponential pulse function \cite{garcia_auto-correlation_2017,theodorsen_relationship_2017}. If all pulses have the same size and velocity, the average radial plasma profile is exponential with the e-folding length given by the product of the pulse velocity and the parallel transit time \cite{garcia_stochastic_2016,militello_scrape_2016,militello_relation_2016,losada_stochastic_2023,losada_stochastic_2024}.

Here we utilize an improved time-delay estimation (TDE) technique on GPI data to obtain new insights into the turbulence-driven cross-field transport in the SOL of magnetically confined plasmas. We apply the TDE technique for one ohmic plasma and two types of H-mode plasmas, an edge localized mode (ELM) free and an enhanced D-alpha (EDA) H-mode, to obtain velocities of large-amplitude, blob-like structures. In all confinement modes investigated, blob velocities and sizes are shown to be similar. The SOL is dominated by blob-like filaments moving radially outwards with velocities in the range from $400$ to $1000\,\text{m}/\text{s}$. This is in agreement with reports from other devices \cite{antar_turbulence_2008,ionita_radial_2013,fuchert_blob_2014,zweben_blob_2016}, and gives further evidence for universality of the statistical properties of plasma fluctuations in the SOL, as well as their role for enhanced plasma--wall interactions.

\section{Experimental setup}\label{sec:exp}

\begin{figure}[tb]
\centering
\includegraphics[width=5cm]{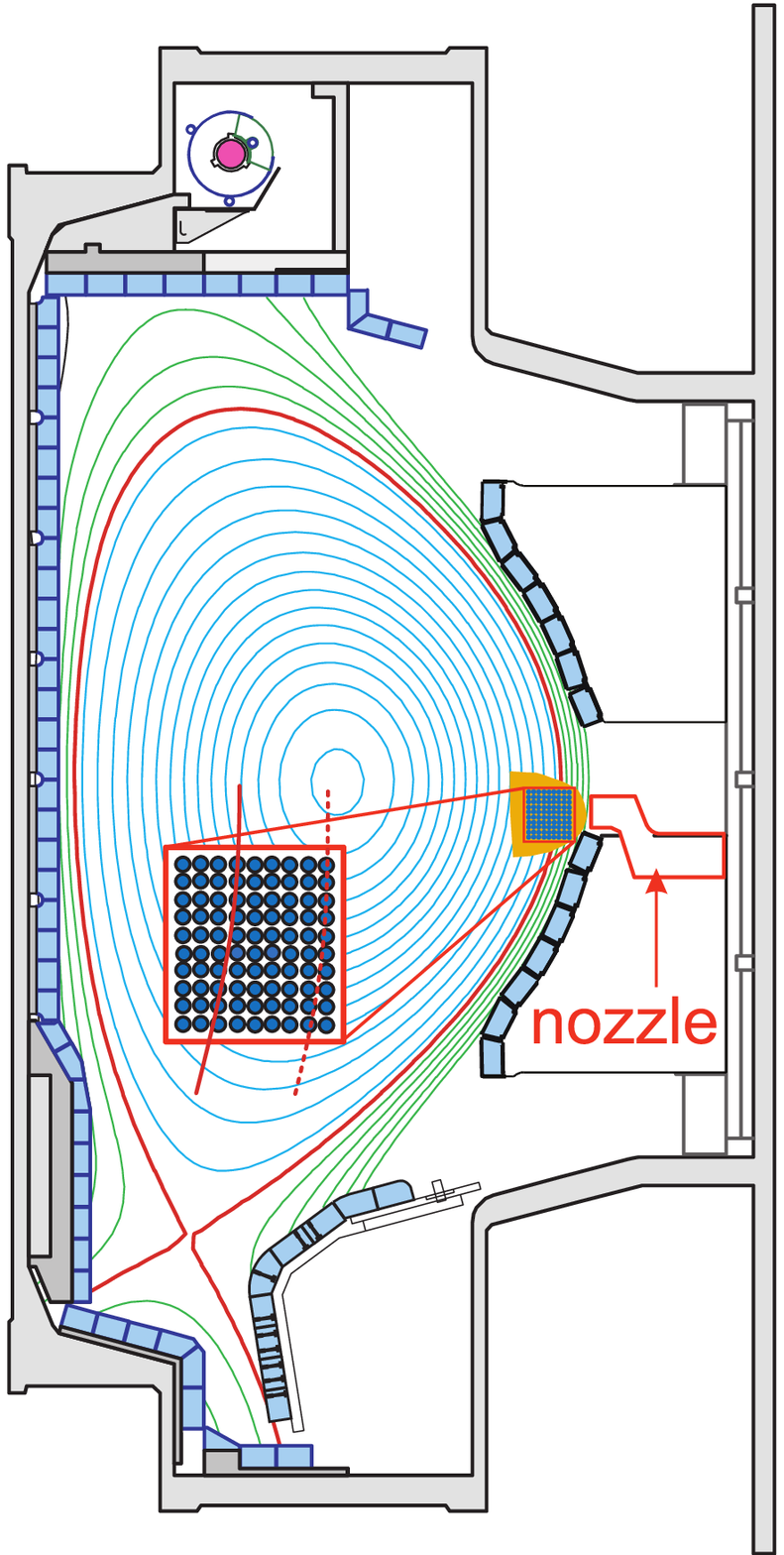}
\caption{Poloidal cross-section of the Alcator C-Mod tokamak showing the GPI diagnostic system comprised by the $9\times10$ avalanche photo diode view channels whose intensity signals are amplified by a neutral gas puff from a nearby nozzle.}
\label{fig:cmod}
\end{figure}

The Alcator C-Mod device \cite{greenwald_20_2014} is equipped with an avalanche photo diode gas puff imaging (GPI) system, illustrated in \Figref{fig:cmod}. This consists of a $9\times10$ array of in-vessel optical fibres with toroidally viewing, horizontal lines of sight \cite{cziegler_experimental_2010,cziegler_ion-cyclotron_2012,zweben_invited_2017}. The plasma emission collected in the views is filtered for He I ($587\,\text{nm}$) line emission that is locally enhanced in the object plane by an extended He gas puff from a nearby nozzle. Because the helium neutral density changes relatively slowly in space and time, rapid fluctuations in He I emission are caused by local plasma density and temperature fluctuations.

The optical fibers are coupled to high sensitivity avalanche photo diodes and the signals are digitized at a rate of $2\times10^6$ frames per second. The viewing area covers the major radius $R$ from $88.00$ to $91.08\,\text{cm}$ and vertical coordinate $Z$ from $-4.51$ to $-1.08\,\text{cm}$ with an in-focus spot size of $3.8\,\text{mm}$ for each of the 90 individual channels. The radial position of the last closed flux surface at the vertical position $Z=-2.99\,\text{cm}$ is according to magnetic equilibrium reconstruction in the range from $88.8$ to $89.6\,\text{cm}$ for the discharges presented here. The limiter radius mapped to the GPI view position is at $R=91.0\,\text{cm}$ for this vertical position.

We present analysis of GPI measurement data from three different confinement modes on Alcator C-Mod, listed in Table~\ref{tab:plasma}. All of these are deuterium fuelled, lower single null diverted plasmas with an axial magnetic field of $5.4\,\text{T}$. The first is an ohmically heated plasma with line-averaged density $\overline{n}_\text{e}=0.47$ and plasma current $I_\text{p}=0.52\,\text{MA}$. This results in a Greenwald fraction of $f_\text{GW}=\overline{n}_\text{e}/n_\text{GW}=0.45$, where the Greenwald density is given by $n_\text{G}=(I_\text{p}/\pi a^2)10^{20}\,\text{m}^{-3}$ with the plasma current $I_\text{p}$ in units of MA and the minor radius $a$ in units of meters \cite{greenwald_density_2002}.

This ohmically heated plasma is compared to two H-modes with ion cyclotron range of frequency (ICRF) heating; an enhanced D-alpha (EDA) H-mode and an ELM-free H-mode. For these two H-modes, a sheared plasma rotation in the edge region results in a strong particle and heat transport barrier. However, neither of these have ELM relaxations of the edge pedestal. In the EDA H-mode, a quasi-coherent mode (QCM) is observed in the particle density and magnetic fluctuations in the edge region at frequencies between $50$ and $100\,\text{kHz}$ \cite{garcia_intermittent_2018}. The QCM results in enhanced particle transport, preventing impurities from accumulating in the core which allows a steady mode of operation \cite{labombard_new_2014,greenwald_20_2014,labombard_evidence_2005}.

In the ELM-free H-mode, lack of macroscopic instabilities of the edge pedestal results in an accumulation of impurities in the core, which eventually causes a radiative collapse of the plasma. Both the plasma and impurity densities increase monotonically during these ELM-free H-modes \cite{greenwald_20_2014,labombard_evidence_2005}. This confinement mode is therefore inherently transient in nature. It is nevertheless an interesting mode of confinement for investigating far-SOL transport properties in the absence of a transport regulator in the edge region \cite{garcia_intermittent_2018}.

The H-mode datasets analyzed here are the same as analyzed in two previous publications \cite{garcia_intermittent_2018,theodorsen_universality_2018}, carefully chosen such that the GPI measurements are not influenced by the strong electric fields from the ICRF wave antennas. The plasma parameters for the discharges analyzed are presented in Table~\ref{tab:plasma}. All these discharges have an axial magnetic field of $5.4\,\text{T}$ and a minor radius of $0.22\,\text{m}$.

\begin{table*}[tb]
\small
\centering
\begin{tabular}{|c|c|c|c|c|c|c|c|c|c|}
\hline
Shot number & Mode & $P\,/\,\text{MW}$ &  $I_\text{p}\,/\,\text{MA}$ & $\overline{n}_\text{e}\,/\,10^{20}\,\text{m}^{-3}$ & $\overline{n}_\text{e}/n_\text{G}$ & $t_\text{start}$\,/\,s & $t_\text{end}$\,/\,s & $T$\,/\,s & Marker
\\ 
\hline
1110201016 & EDA H & 3.0 & 0.90 & 3.69 & 0.64 & 1.06 & 1.345 & 0.285 & \color{my_magenta_color}$\blacktriangle$
\\
1110201011 & ELM-free H & 3.0 & 1.20 & 4.32 & 0.56 & 1.06 & 1.255 & 0.195 & \color{my_pink_color}\large{${\bullet}$}
\\
1160616018 & Ohmic & 0.0 & 0.52 & 1.65 & 0.47 & 1.15 & 1.450& 0.300 & \color{my_yellow_color}$\blacktriangledown$
\\
\hline
\end{tabular}
\caption{List of plasma discharges giving the shot number, confinement mode, heating power $P$, plasma current $I_\text{p}$, line-averaged density $\overline{n}_\text{e}$, Greenwald density fraction $f_\text{GW}$, start and end time for data analysis, duration $T$ of data time series, and plot marker symbols and colors used in the following figures.}
\label{tab:plasma}
\end{table*}

Figure~\ref{fig:gpiraw} shows the normalized GPI data time series recorded during the EDA H-mode state at various radial positions in the SOL region for $Z=-2.99\,\text{cm}$. Here and in the following, the raw measurement data time series $\Phi(R,Z,t)$ -- the brightness of the He I emission -- are normalized by subtracting a running mean and dividing by a running standard deviation, $\Phitilde=(\Phi-\Phiave)/\Phirms$, using a one millisecond window in order to remove low-frequency trends. Several large-amplitude fluctuation events corresponding to blob-like structures moving radially outwards are observed in \Figref{fig:gpiraw}. The horizontal dotted lines indicate fluctuation level crossings of $2.5$ times the root mean square value above the mean, used as a conditional averaging threshold in the following analysis.

\begin{figure}[tb]
\centering
\includegraphics[width=0.5\textwidth]{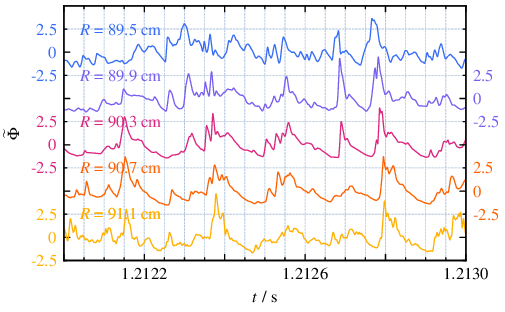}
\caption{GPI measurement time series for the EDA H-mode at various radial positions for $Z=-2.99\,\text{cm}$.}
\label{fig:gpiraw}
\end{figure}

\section{Fluctuation statistics}\label{sec:fluc}

\begin{figure*}[tb]
\centering
\includegraphics[width=0.95\textwidth]{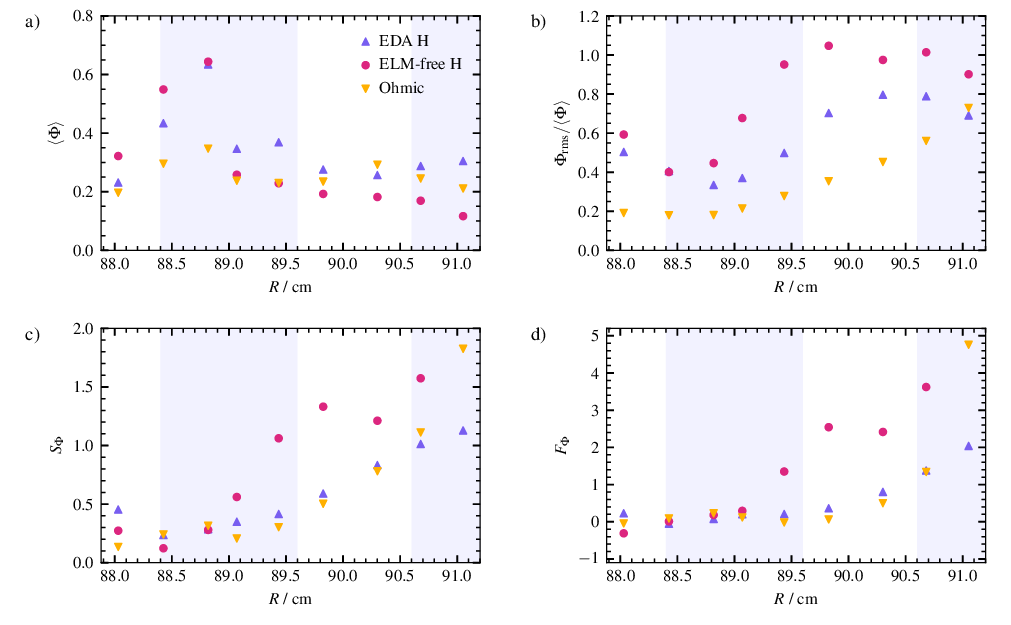}
\caption{Radial variation of vertically averaged (a) average line intensity $\Phiave$, (b) relative fluctuation level $\Phirms/\Phiave$, (c) skewness moment $S_\Phi$, and (d) flatness moment $F_\Phi$ for the various confinement modes.}
\label{fig:gpimoments}
\end{figure*}

The radial variation of the lowest order statistical moments of the GPI line intensity measurements are presented in \Figref{fig:gpimoments}. These are averaged over all vertical diode view positions for each radius. The gray shaded region to the right indicates the limiter shadow, where the magnetic connection length ranges from $10$ to $80\,\text{cm}$ for all confinement modes investigated. In the main SOL the magnetic connection length is approximately $5$ meters for the two H-mode discharges, and approximately $8$ meters for the Ohmic plasma state. The gray shaded region to the left indicates the location of the last closed magnetic flux surface, which varies vertically as well as in time during each discharge.

The moments include the mean value $\Phiave$, the relative fluctuation level $\Phirms/\Phiave$, the skewness moment $S_\Phi$ and the flatness moment $F_\Phi$. At the outermost diode view position, the skewness is $2.5$ and flatness is $14$ for the ELM-free H-mode. These are not included with the plot range used. Figure~\ref{fig:gpimoments} shows that the fluctuation level as well as the intermittency of the fluctuations increase radially outwards in the SOL. The skewness and flatness moments are comparable for the Ohmic and EDA H-modes, while the ELM-free H-mode has higher intermittency.

\begin{figure}[tb]
\centering
\includegraphics[width=0.5\textwidth]{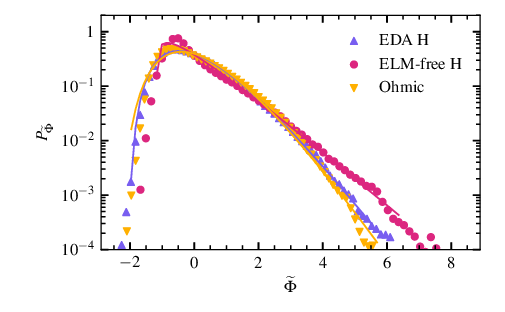}
\caption{Probability density function of the normalized GPI line intensity fluctuations at $(R,Z)=(90.69,-2.99)\,\text{cm}$. The solid curves show the best tail fits of a Gamma distribution for the various confinement modes.}
\label{fig:gpipdf}
\end{figure}

The probability distributions $P_{\Phitilde}$ for the normalized line intensity fluctuations in the far-SOL at $(R,Z)=(90.69,-2.99)\,\text{cm}$ are presented in \Figref{fig:gpipdf}. The solid curves show the best tail fits of a Gamma distribution for the various confinement modes yielding a shape parameter $\gamma=6.2$ for the Ohmic plasma state, $\gamma=1.4$ for the ELM-free H-mode, and $\gamma=4.2$ for the EDA H-mode. This is consistent with the skewness and flatness moments presented in \Figref{fig:gpimoments}.

\begin{figure}[tb]
\centering
\includegraphics[width=0.5\textwidth]{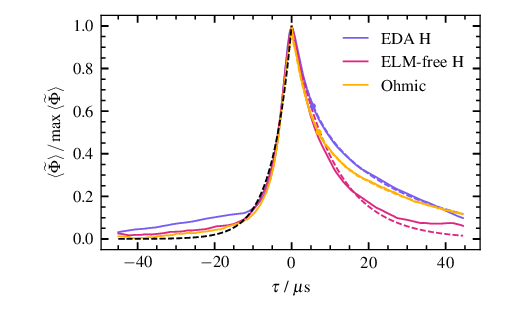}
\caption{Conditionally averaged burst shape of normalized GPI line intensity fluctuations at $(R,Z)=(90.69,-2.99)\,\text{cm}$ for the various confinement modes. The dashed lines for positive lags are the best fit of a bi-exponential function with the filled circles indicating the transition between the two decay regions.}
\label{fig:gpica}
\end{figure}

A standard conditional averaging method has been used to identify the characteristic shape of large-amplitude bursts in the measurement time series. Using a threshold condition of $2.5$ times the standard deviation above the mean value, $\text{max}\,\Phitilde>2.5$, gives conditionally averaged pulse shapes. This is presented in \Figref{fig:gpica} for the three confinement modes investigated, at the location $(R,Z)=(90.69,-2.99)\,\text{cm}$ in the far-SOL. The fast rise of these bursts are well described by an exponential function with a rise time of approximately $5\,{\mu}\text{s}$. The following pulse decay is for the ELM-free H-mode well described by an exponential function with a decay time of $11\,\mu\text{s}$, while for the ohmic plasma state and the EDA H-mode, the pulse decay is well described by a bi-exponential function. The ohmic plasma state has a fast transient decay time of $4.3\,\mu\text{s}$ and a slower asymptotic decay time of $30\,\mu\text{s}$. For the EDA H-mode these decay times are $4.6\,\mu\text{s}$ and $26\,\mu\text{s}$, respectively.

\begin{figure}[tb]
\centering
\includegraphics[width=0.5\textwidth]{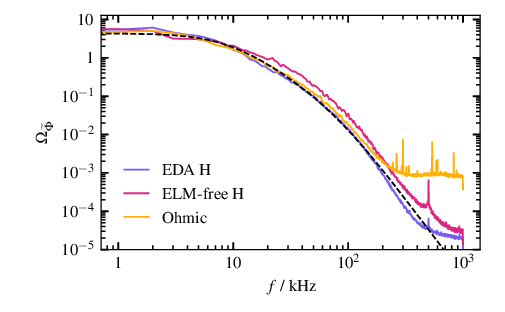}
\caption{Frequency power spectral density of normalized GPI line radiation fluctuations at $(R,Z)=(90.69,-2.99)\,\text{cm}$. The dashed line shows the spectrum for a two-sided exponential pulse function with duration $20\,\mu\text{s}$ and asymmetry parameter $1/10$.}
\label{fig:gpipsd}
\end{figure}

The frequency power spectral densities for the GPI fluctuations in the far-SOL at $(R,Z)=(90.69,-2.99)\,\text{cm}$ are presented in \Figref{fig:gpipsd}. The spectra are nearly identical for the various confinement modes, consistent with the similarity of the conditional burst shapes in \Figref{fig:gpica}. They are well described by the spectrum of a two-sided exponential pulse function with a duration of $20\,\mu\text{s}$ and asymmetry parameter $1/10$ \cite{garcia_intermittent_2018}. The longer duration estimated from conditional averaging is likely due to overlap of pulses.

\begin{figure}[tb]
\centering
\includegraphics[width=0.5\textwidth]{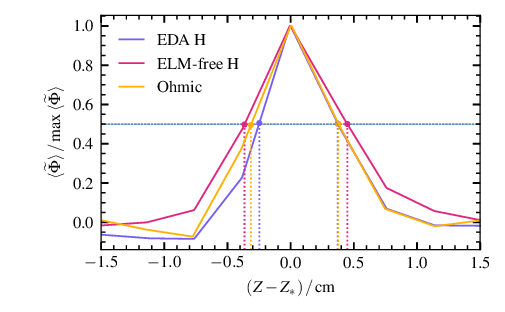}
\caption{Vertical variation of large-amplitude blob structures obtained by conditional averaging for various confinement modes.}
\label{fig:gpiphiz}
\end{figure}

By using the conditional averaging method, we can also obtain an estimate of the vertical size of large-amplitude blob structures. At a given reference position $(R_*,Z_*)$, we identify conditional events that have a local maximum in time, $\text{max}\,\Phitilde(R_*,Z_*)>2.5$, as well as a vertical maximum at this position, $\Phitilde(R_*,Z_*)>\Phitilde(R_*,Z\lessgtr Z_*)$. The resulting vertical blob shapes for the various confinement modes are presented in \Figref{fig:gpiphiz} at the reference position $(R_*,Z_*)=(90.30,-3.00)\,\text{cm}$. This reveals a full width at half maximum of $\ell=0.62\,\text{cm}$ for the EDA H-mode, $\ell=0.81\,\text{cm}$ for the ELM-free H mode, and $\ell=0.69\,\text{cm}$ for the ohmic plasma state. Thus, the vertical blob size is comparable for all confinement modes.

Previous investigations have shown that peak amplitudes and waiting times between large-amplitude bursts are both exponentially distributed for these ohmic and H-mode plasmas \cite{garcia_intermittent_2018,theodorsen_universality_2018}. The statistical properties of the fluctuations presented here are consistent with the stochastic model describing the fluctuations as a super-position of uncorrelated pulses, satisfying the underlying assumptions of the model as well as its predictions.

\section{Time-delay estimation of blob velocities}

\begin{figure*}[tb]
\centering
\includegraphics[width=0.33\textwidth]{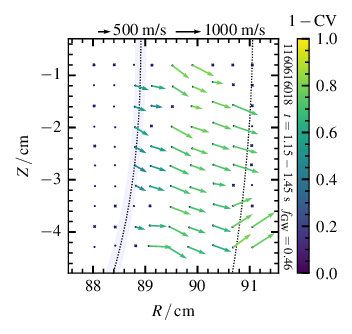}
\includegraphics[width=0.33\textwidth]{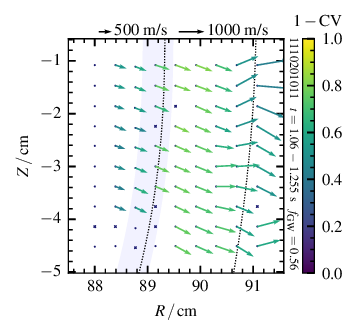}
\includegraphics[width=0.33\textwidth]{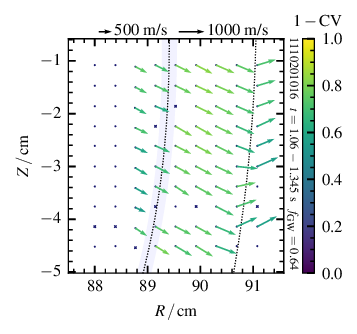}
\caption{Time-delay estimation of blob velocities for ohmic plasma (left), ELM-free H-mode (center) and EDA H-mode (right) based on cross-conditional averaging with nearest neighbor diode view positions. In each plot, the gray shaded region to the left is the location of the separatrix which varies in time, and the dotted line to the right is the limiter structure. The color coding gives conditional reproducibility.}
\label{fig:gpitde}
\end{figure*}

An improved time-delay estimation (TDE) method for blob velocities has recently been developed for coarse grained imaging data \cite{losada_three-point_2024,losada_time_2024}. Using three non-aligned measurement points, this method reliably estimates both radial and poloidal components of the velocity vector associated with blob-like structures moving across the magnetic field lines towards the main chamber wall. In \Figref{fig:gpitde} we present the result from this TDE analysis for the three magnetic confinement modes on Alcator C-Mod. The GPI measurement points marked by a cross are broken diode view channels for which no measurement data exists.

Each velocity vector is obtained from TDE, calculated from all its neighboring GPI diode view positions; above, below, to the left and to the right. A cross-conditional averaging method is used to estimate the time delay between measurement points, based on peak amplitudes larger than 2.5 times the standard deviation above the mean value, $\text{max}\,\Phitilde>2.5$. Thus, the estimated velocity components are associated with large-amplitude fluctuations, specifically the $2.5$ level threshold crossing seen in \Figref{fig:gpiraw}. The color bar in \Figref{fig:gpitde} gives the conditional reproducibility $1-\text{CV}$, where $\text{CV}$ denotes the conditional variance \cite{kube_fluctuation_2016}.

For each diode view position, a velocity vector is estimated based on the following criteria: (i) measurement data exists for a least one horizontally and one vertically separated diode view channel, (ii) the cross-conditionally averaged pulse function must be unimodal, (iii) the time lag for maximum correlation between either the horizontally or the vertically separated measurement points must be larger than the sampling time, (iv) the fluctuations are not influenced by strong electric fields due to mapping along magnetic field lines to the ICRF antennas, and (v) the fluctuation statistics must be consistent with blob dynamics, in particular a Lorentzian-shaped frequency power spectral density such as that presented in \Figref{fig:gpipsd}. No velocity vector is assigned unless all these criteria are met.

\begin{figure}[tb]
\centering
\includegraphics[width=0.5\textwidth]{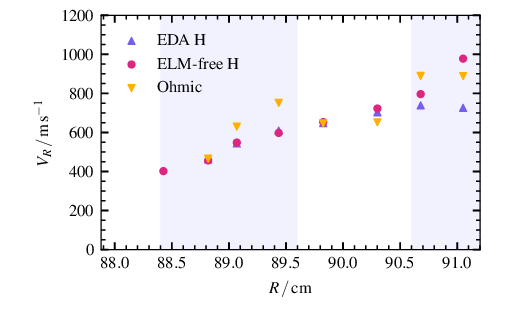}
\caption{Radial component of conditional time-delay estimated velocities averaged over each vertical column for various confinement modes.}
\label{fig:gpitdezave}
\end{figure}

The vector plots in \Figref{fig:gpitde} show that the entire SOL region, from the separatrix and outwards, is dominated by radial motion of blob-like structures. This is consistent for all confinement modes investigated. In the main SOL, the conditional reproducibility is higher than $50\%$. The radial velocity components are averaged over all vertical diode view positions for each radius, resulting in an average radial velocity profile presented in \Figref{fig:gpitdezave}. The magnitude of the radial velocity components increases from approximately $400\,\text{m}/\text{s}$ at the separatrix up to nearly $1000\,\text{m}/\text{s}$ in the limiter shadow for all three confinement modes.  Averaging the radial velocity component over all diode view positions results in a SOL-averaged velocity of $703\,\text{m}/\text{s}$ for the ohmic plasma, $644\,\text{m}/\text{s}$ for the ELM-free H-mode, and $632\,\text{m}/\text{s}$ for the EDA H-mode. Similar results are obtained by using cross-correlation functions rather than cross-conditional averaging \cite{losada_three-point_2024}.

\section{Discussion and conclusions}

The results presented here provides further evidence for universality of the statistical properties of plasma fluctuations in the SOL. In all confinement modes investigated -- ohmic plasma, ELM-free H-mode and EDA H-mode -- the radial motion of blob structures results in large-amplitude fluctuations and high average plasma density. In the far SOL, these fluctuations are well described as a super-position of uncorrelated pulses, which is confirmed by the two-sided exponential burst shapes seen in \Figref{fig:gpica}, the Gamma PDFs in \Figref{fig:gpipdf} and the Lorentizan-shaped frequency PSDs in \Figref{fig:gpipsd}. These statistical properties show similarities across the three confinement modes.

An improved TDE method has been utilized to obtain velocities of blob-like structures, which also reveals similarities between ohmic and H-mode plasmas. Here, radial velocity components range between $400$ and $1000\,\text{m}/\text{s}$ for the three confiment modes, as seen in \Figref{fig:gpitdezave}. Although the radial blob velocities from the TDE analysis increases radially outwards, this does not necessarily imply that there is a corresponding acceleration of individual structures. Their dynamic is rather complicated, with randomness in the location at which they are formed. Small-amplitude blobs can also stagnate during their radial motion, as seen in \Figref{fig:gpiraw}.

The radial transit time for propagating blobs is given as their size divided by their velocity, $\tau_\perp=\ell/v$. Since conditional averaging and TDE reveal that blob sizes and velocities are comparable for all confinement modes investigated, the radial transit time will be similar for all of them as well. This is consistent with the close similarity of the conditionally averaged burst shapes in \Figref{fig:gpica} and the frequency spectra presented in \Figref{fig:gpipsd}. We note that since the velocities presented in \Figref{fig:gpitde} show dominating radial components for each confinement mode, the vertical variation in \Figref{fig:gpiphiz} can be used accurately to estimate blob sizes. The stochastic model gives a shape parameter of the PDFs in \Figref{fig:gpipdf} that is given by the ratio of the average pulse duration and waiting times $\gamma=\tau_\text{d}/\tau_\text{w}$ \cite{garcia_stochastic_2012}. Since the radial transit time dominates the pulse duration, we conclude that the difference between the various confinement modes is largely due to changes in the average waiting time between large-amplitude blob structures.

Stochastic modeling of intermittent fluctuations in the SOL predicts an exponential radial plasma density profile with an e-folding length given by the product of the radial blob velocity and the parallel transit time \cite{garcia_stochastic_2016,militello_scrape_2016,militello_relation_2016,losada_stochastic_2023,losada_stochastic_2024}. The comparable blob velocities in ohmic and high confinement modes, as seen in \Figref{fig:gpitdezave}, explains the similar broad and flat profiles to be expected in ELM-free and EDA H-modes as those observed in ohmic plasmas \cite{labombard_new_2014,labombard_evidence_2005}.

\section*{Acknowledgements}

This work was supported by the UiT Aurora Centre Program, UiT The Arctic University of Norway (2020). Alcator C-Mod measurement data was generated under the US DoE award DE-FC02-99ER54512. A.~D.~H.\ and O.~E.~G.\ acknowledge the generous hospitality of the MIT Plasma Science and Fusion Center where this work was conducted.

\bibliographystyle{elsarticle-num} 
\bibliography{SOL}

\end{document}